# Estimation of Participation Factors for Power System Oscillation from Measurements

Tianwei Xia, *Member, IEEE*, Zhe Yu, *Member, IEEE*, Kai Sun, *Fellow, IEEE*, Di Shi, *Member, IEEE,* Kaiyang Huang, *Student Member, IEEE*

*Abstract*—In a power system, when the participation factors of generators are computed to rank their participations into an oscillatory mode, a model-based approach is conventionally used on the linearized system model by means of the corresponding right and left eigenvectors. This paper proposes a new approach for estimating participation factors directly from measurement data on generator responses under selected disturbances. The approach computes extended participation factors that coincide with accurate model-based participation factors when the measured responses satisfy an ideally symmetric condition. This paper relaxes this symmetric condition with the original measurement space by identifying and utilizing a coordinate transformation to a new space optimally recovering the symmetry. Thus, the optimal estimates of participation factors solely from measurements are achieved, and the accuracy and influencing factors are discussed. The proposed approach is first demonstrated in detail on a two-area system and then tested on an NPCC 48-machine power system. The penetration of inverter-based resources is also considered.

## I. BACKGROUND

IN an interconnected power system, rotor angle oscillations among generators are prevalent, as evidenced by real-time wide-area measurements under both ambient and contingency conditions. Such oscillations, especially low-frequency inter-area oscillations, can affect the safety margins for grid operations [1] and cause stability issues of generators [2]. The Participation Factor (PF), as a useful small-signal analysis tool, can evaluate how each state variable of a generator or dynamic device participates in an oscillatory mode [3]. PFs of state variables regarding a specific oscillatory mode, i.e., a pair of complex eigenvalues, are conventionally calculated as the products of corresponding elements from both the right and left eigenvectors of the system's linearized model. Comparatively, the mode shape and mode composition as another two modal properties focus only on one-way linkages between state variables and the mode by utilizing either the right or left eigenvector. Thus, a PF captures both the response of a state variable with the mode, i.e., the mode shape, and its contribution to the mode, i.e., the mode composition, evaluating a bidirectional connection between the state variable and mode.

Accurate estimation of PFs can largely benefit grid operations in the real-time environment. For instance, when an oscillation occurs, knowing how generators participate in the oscillation mode can pinpoint the most involved or susceptible generators for early mitigation before instability is caused. The generators with high PFs can be the candidate locations to perform damping control using Power System Stabilizers (PSS) [3]. Recently, PFs have also been applied to renewable energy systems for the identification of crucial state variables in phase-locked loops [4-6] and DC links [7] associated with sub-synchronous oscillations. Furthermore, PFs are used to associate oscillation damping with generator outputs for strategic generation re-dispatches [8][9].

For practical applications, variant linear PFs have been proposed to satisfy different application demands. For instance, reference [10] introduces loop participation factors that evaluate the influences of individual components on system modes by leveraging the loop and nodal observability and controllability. Reference [11] proposes impedance participation factors to analyze the sensitivity of black box models, defined in terms of the residue of the whole-system admittance matrix with the chain rule. A similar impedance participation factor is discussed in [12], focusing on frequency domain performance. In addition, a recent study in [13] advances modal PFs for exploring the role of algebraic variables in modes more comprehensively.

Alternative methodologies and definitions for PFs have been explored extensively in literature [14]-[22], where the classic PF concept based on the state matrix has been expanded through perspectives of probability and nonlinearity. Reference [17] broadens the PF definition to correlate with the initial system state, enabling PFs to represent an average connection between a state variable and a mode across responses from various initial states. This "extended PF" has been validated against traditional model-based PFs, contingent on the distribution of initial states fulfilling an ideally symmetric condition. The advantages of these extended PFs in power systems are illustrated in [18] and [19]. References [20] and [21] distinguish between mode-in-state PFs and state-in-mode PFs, using a model-based calculation approach that relies on eigenanalysis on the system's linearized model. Another study [22] examines the uniqueness of the extended PF and introduces the energy-based PF as an alternative.

Tianwei Xia, Kai Sun, and Kaiyang Huang are with the Electrical Engineer and Computer Science Department, University of Tennessee, Knoxville, TN 37996 USA (e-mail: tianweixia@gmail.com, kaisun@utk.edu, khuang12@vols.utk.edu).

Z. Yu is with Didi Chuxing (e-mail: yzae2623@gmail.com).

D. Shi is with the Electrical and Computer Engineering Department, New Mexico State University, Las Cruces, NM 88003 USA (e-mail: dshi@nmsu.edu).

Conventionally, PFs are calculated from a system's linearized model. Most existing methods presume access to all right (or left) eigenvectors of the linearized model. However, a linearized model for the whole power grid is not always available in practice, especially with high inverter-based resource penetration [23]. An alternative method involves using wide-area measurements to monitor the responses of generators and other dynamic devices to small disturbances, thus enabling the estimation of their PFs directly from measurement data. However, there is very limited research on measurement-based PF estimation for power systems. Reference [24] employs a Koopman operator-based approach for measurement-based PFs and allows for estimating both linear and nonlinear PFs within the Koopman mode. This approach relies on the choice of the observable function, which is a question still unresolved in the field [25]. Reference [26] also determines PFs from measurement data, where the PFs are estimated by certain perturbation on the desired variables; however, these response-based PFs require a black-box model and carefully designed system perturbation for simulations, significantly limiting their practical use.

This paper is dedicated to estimating the linear PFs of generators within oscillatory modes directly from measurement data and proposes a practical methodology. The key contributions of the paper include: 1) a systematic approach to computing PFs from measurement data, which uses extended PFs as a bridge but relaxes the symmetric condition by finding a coordinate transformation; 2) a PF estimation methodology applicable to black-box models by generalizing response-based PFs in [26]; 3) demonstration and validation of the new approach on a large-scale model considering the penetration of inverter-based resources (IBRs).

The rest of the paper is organized as follows: Section II details the proposed measurement-based approach, including the determination of coordinate transformations for optimal state space symmetry, the computation of extended PFs, and the method of translating these calculated PFs back to the original state space. In Section III, the proposed method is initially applied to a two-area system and later tested on a 48-machine Northeastern Power Coordinating Council (NPCC) model. Conclusions are drawn in Section IV.

## II. Proposed Measurement-based Method For Estimation of PF

### A. Preliminary

Consider an $N$-dimensional linear system in (1) describing a power system model linearized at its stable equilibrium:

$$\dot{\mathbf{x}} = \mathbf{A}\,\mathbf{x}\;. \qquad (1)$$

Let $\lambda_i$ denote the $i$-th eigenvalue of matrix $\mathbf{A}$. Its left eigenvector $\mathbf{\psi}_i = [\psi_{i1, \dots} \psi_{iN}]$ and right eigenvector $\mathbf{\phi}_i = [\phi_{1i, \dots} \phi_{Ni}]^{\mathrm{T}}$ respectively define the composition and shape of the mode. Note that if $\lambda_i$ is complex, it and its conjugate together define one oscillatory mode of matrix $\mathbf{A}$, referred to as mode $i$. Since the eigenvectors are also conjugates and convey identical information, modal properties can be investigated based on

either $\lambda_i$ or its conjugate. Then, the PF of the $k$-th state in mode $i$ is defined by

$$\mathrm{PF}_{ki} \overset{\mathrm{def}}{=} p_{ki} = \psi_{ik}\phi_{ki}\;, \qquad (2)$$

as a dimensionless value. Notice that (2) provides a general model-based approach to calculate PF from the shape $\phi_{ki}$ and composition $\psi_{ik}$ of the mode.

When an accurate matrix $\mathbf{A}$ is unavailable, PFs can still be estimated as follows. The response of the $k$-th state variable subject to a small disturbance can be written as a combination of $N$ modes

$$x_k(t) = \sum_{i=1}^{N} B_{ki} e^{\lambda_i t} = \sum_{i=1}^{N} (\mathbf{\psi}_i \mathbf{x}_0)\phi_{ki} e^{\lambda_i t} \quad (k=1,\dots,N), \qquad (3)$$

where $\mathbf{x}_0 = [x_{01}, \dots, x_{0N}]^{\mathrm{T}}$ is the initial state vector, and modal amplitude $B_{ki} = \mathbf{\psi}_i \mathbf{x}_0 \phi_{ki}$ is proportional to the mode shape $\phi_{ki}$, which evaluates how much mode $i$ is excited in the $k$-th state variable. From state responses, $\phi_{ki}$ can be estimated by normalizing amplitude $B_{ki}$ with respect to the initial condition $\mathbf{x}_0$ [27]. Accordingly, paper [17] proposes Extended Participation Factors (EPFs) defined by (4), utilizing all state responses starting from a set of initial states:

$$\mathrm{EPF}_{ki} \overset{\mathrm{def}}{=} \underset{x_{k0}^{(l)} \in S}{avg}\, \frac{B_{ki}^{(l)}}{x_{k0}^{(l)}}, \qquad (4)$$

where $S$ is the compact set of the initial values of the, and "($l$)" indicates the element index of the set $S$. It has been proved that EPFs are equivalent to conventional participation factors when initial states satisfy a symmetric condition (defined below). Since $B_{ki}$ can be derived from measured responses via data-driven modal analysis, this extension facilitates the direct estimation of PFs from measurement data. To avoid any confusion, (4) is designated as the EPF for clarity, following the terminology used in references [18] and [19]. Notably, this term is referred to by a different name in reference [17].

A *symmetric condition* in [17] refers to the symmetry of the initial state set $S$ in the view of any dimension. In other words, for any $k \in \{1, 2, \dots, N\}$, any state $z = (z_1, \dots, z_k, \dots, z_N) \in S$ implies $(z_1, \dots, -z_k, \dots, z_N) \in S$. Suppose the initial state set $S$ satisfies this symmetric condition in the state space. In that case, the state variables' distribution exhibits independence, and then the EPFs are equal to conventional PFs calculated from eigenvectors. This can be proved as:

$$\mathrm{EPF}_{ki} = \psi_{ik}\phi_{ki} + \frac{1}{L}\sum_{l=1}^{L}\sum_{j=1,\,j\neq k}^{N} \psi_{ji}\phi_{ki}\frac{x_{j0}^{(l)}}{x_{k0}^{(l)}} = \psi_{ik}\phi_{ki} = \mathrm{PF}_{ki}, \quad (5)$$

where $N$ is the system dimension, and $L$ is the total number of the initial state set. Variable $x^{(l)}_{j0}$ denotes the $j$-th state for element $l$. Consider a 2-dimensional system where the four initial states form a rectangle in the state space, representing ideal symmetry (as shown in the bottom right figure of Fig. 1). Under this condition, the EPFs are the same as the PFs.

### B. The main idea of the new approach

However, when a power system experiences oscillations subjected to a disturbance, the collected field measurements may not necessarily satisfy this symmetric condition. This paper introduces a new measurement-based approach for calculating PFs even when the symmetry condition is not satisfied, as illustrated in Fig. 1.

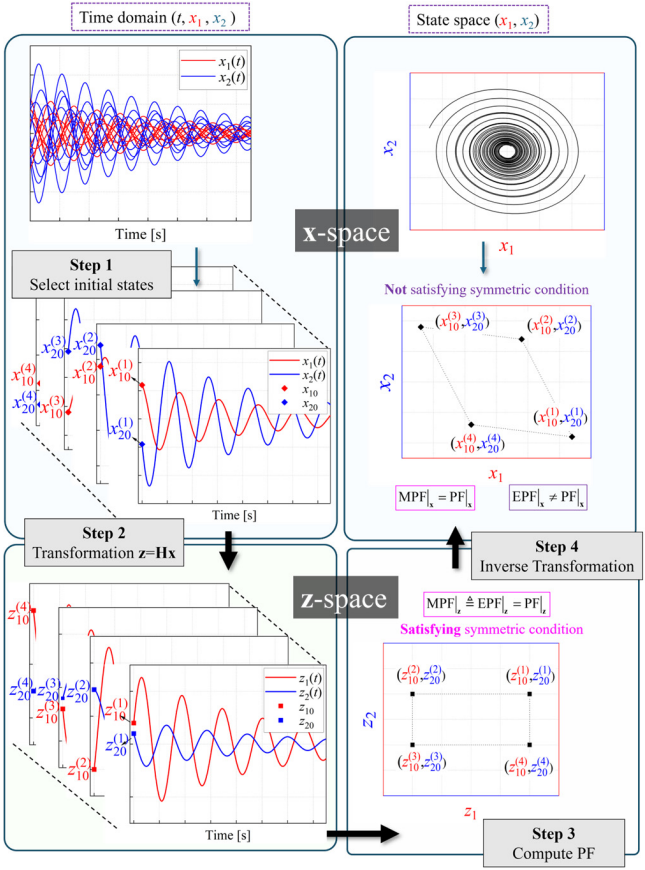

**Fig. 1.** Structure of new measurement-based approach. (**PF**: Participation factors, **MPF**: Measurement-based PF, **EPF**: Extended PF)

The idea of the new approach is to introduce a linear transformation **H**, which maps the original state space on **x** into a new state space on **z**. It is expected that in this transformed **z**-space, selected measurement data can exhibit the most symmetric distribution, enabling the EPFs computed using (4) to best approximate the true PFs in **z**-space. Subsequently, the calculated EPFs in **z**-space are transformed back to the original **x**-space, ultimately providing estimates of PFs. In this paper, the PFs estimated from measurements using the proposed method are referred to as Measurement-based Participation Factors or MPFs for brevity.

To avoid ambiguity, TABLE I presents detailed information about the three types of PFs studied in this paper. In the rest of Section II, the four steps of the proposed approach are detailed, followed by a discussion on errors in PF estimation.

TABLE I THE DIFFERENCE BETWEEN DIFFERENT KINDS OF PFs

| Name | Formula | Data Source | Description |
|---|---|---|---|
| PF | (2) | Model-based | Based on the PF definition |
| EPF | (4) | Model or measurement-based | Equal to PF under the symmetric condition |
| MPF | (3) | Measurement-based | Applying EPF as the bridge for the optimal estimate of PF from measurements |

The steps are as follows:

**Step 1**: Select the optimal set of initial states from measurements.

**Step 2**: Find the transformation **H** toward a **z**-space to best meet the symmetric condition.

**Step 3**: Compute the MPFs in **z**-space.

**Step 4**: Translate the MPFs back to **x**-space.

*1) Step 1 Selecting the optimal set of initial states*

In $D$-dimensional state space, a set of initial states satisfying the symmetric condition has a symmetric distribution and appears as symmetrical pairs around the equilibrium, i.e., the origin in (1). Assume that the initial states associated with $2^D$ selected measurement segments can form all vertices of a $D$-dimensional parallelotope in the **x**-space, as illustrated in Fig. 3 for $D = 2$ by the parallelograms and rectangles. Then, after the linear transformation, these initial states form a hyperrectangle in the **z**-space. However, from real-world measurements, it might not always be feasible to find an ideal $D$-dimensional parallelotope. Thus, the most symmetric set of initial states can be identified by solving this optimization problem:

$$\min_{\mathbf{x}_{A'}} \lVert \mathbf{x}_A + \mathbf{x}_{A'} \rVert \quad s.t. \; \lVert \mathbf{x}_A \rVert, \lVert \mathbf{x}_{A'} \rVert > r_{threshold} \,, \quad (6)$$

where the norm can take the Euclidian distance when state variables are considered to be the same type, $\mathbf{x}_A$ is the state vector in the set of initial states, and $\mathbf{x}_{A'}$ is its most symmetric peer in the set. A practical consideration in this context is the robustness against noise in measurements. The farther a measured state is from the origin, the less it is affected by noise. Therefore, it is advisable to select initial states that are not too close to the origin by defining their minimum distance, $r_{threshold}$, to the origin as specified in (6).

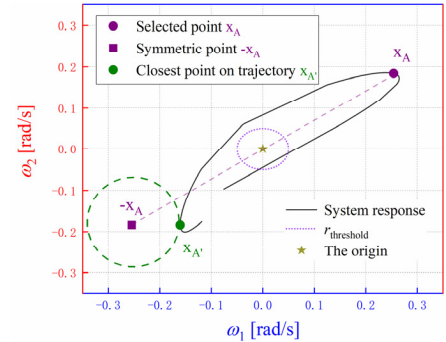

**Fig. 2.** The procedure to find a symmetric pair of initial states.

The process of identifying the most symmetric peer for a selected state is depicted in Fig. 2. Point $\mathbf{x}_A$ is selected along the trajectory, indicated by a purple circle, while its ideally symmetric state, $-\mathbf{x}_A$, is represented by a purple square; however, this point does not lie on the trajectory. Instead, another point, $\mathbf{x}_{A'}$, which is on the trajectory and closest to $-\mathbf{x}_A$ with the minimum distance $\lVert \mathbf{x}_A + \mathbf{x}_{A'} \rVert$, is identified as the solution to the optimization problem mentioned earlier. In this way, all such most symmetric pairs collectively approximate a parallelotope in the **x**-space. During Step 2, this parallelotope is transformed into an approximate hyperrectangle in **z**-space, ensuring the symmetric condition in **z**-space is met. In addition, to enhance the time performance, the KD tree approach [28] is also applied to search for the optimal symmetric pairs.

### 2) Step 2: Finding the transformation

The initial states are selected as symmetric pairs around the origin in the **x**-space, forming the vertices of a parallelotope. The next step involves identifying the transformation **H** that maps this parallelotope to a hyperrectangle. A method to determine the desired transformation is proposed in this step, which is exemplified for a two-dimensional system in Fig. 3.

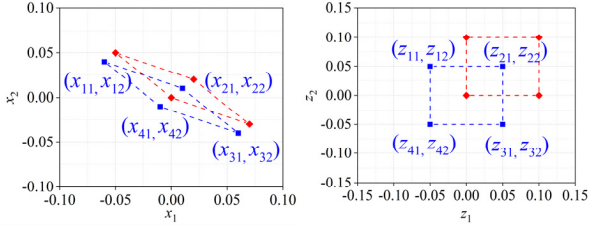

**Fig. 3.** The transformation of an initial state set.
(Left: Initial states in **x**-space. Right: Initial states in **z**-space)

The initial state set optimized in *Step 1* forms the blue parallelogram centered at the origin, as depicted in Fig. 3 (left). This parallelogram can be transformed into the red parallelogram by shifting the center to one of its vertices. Subsequently, a transformation is applied to convert the red parallelogram into the red rectangle shown in Fig. 3 (right), which is then translated back to the original origin. It becomes clear that the transformation from the red parallelogram to the red rectangle can be determined based on the edges of the parallelogram in x-space, as detailed below:

$$\begin{bmatrix} z_{i1} \\ z_{i2} \end{bmatrix} = \begin{bmatrix} x_{11} - x_{41} & x_{31} - x_{41} \\ x_{12} - x_{42} & x_{32} - x_{42} \end{bmatrix}^{-1} \begin{bmatrix} x_{i1} \\ x_{i2} \end{bmatrix} = \mathbf{H} \begin{bmatrix} x_{i1} \\ x_{i2} \end{bmatrix}, \quad (7)$$

where $x_i$ is the initial state set in **x**-space, and $z_i$ is the initial state in **z**-space. The translations to a new origin in **x**-space (from the blue parallelogram to the red parallelogram) and back to the original origin in **z**-space (from the red rectangle to the blue rectangle) effectively cancel each other out. Consequently, the desired transformation from **x**-space to **z**-space (from the blue parallelogram to the blue rectangle) remains unchanged as **H**. In other words, **H** is invariant following a coordinate translation. This observation is valid for an *N*-dimensional system, as demonstrated by Lemma 1 in the Appendix, where detailed proof of this property is also provided. This characteristic can be formally stated as a theorem that a transformation from a parallelotope in **x**-space to a hyperrectangle in **z**-space that are both centered at the origin of the dimension *N* can be determined from any *N*+1 pairs of vertices in **x**-space, denoted by {**x₀**, **x'₀**}, {**x₁**, **x'₁**}, ..., {**xₙ**, **x'ₙ**}, as **H**=[**x₁**-**x₀**, ..., **xₙ**-**x₀**]⁻¹.

### 3) Steps 3 and 4: Computing MPFs in z- and x-spaces

After determining the parameters of the linear transformation in the previous subsection, the trajectory in the **x**-space can also be transformed into the **z**-space. The $B_{ki}$ values in the **z**-space are obtained using Prony's analysis [27]. In **z**-space, the symmetric condition is met, enabling the computation of EPFs using (4). It is important to note that if a black-box model is employed, the first three steps can be ignored, and the MPFs in the **x**-space can be directly calculated. This method serves as an extension of the response-based approach, with details available in [26].

Theoretically, if all modes and their shapes or compositions are known, the calculation of MPFs becomes straightforward, given the established inverse relationship between the matrices made by right and left eigenvectors [29]

$$\mathbf{P} = \mathbf{\Phi} \circ \mathbf{\Psi}^{\mathrm{T}} = \mathbf{\Phi} \circ \mathbf{\Phi}^{-\mathrm{T}}, \quad (8)$$

where "∘" is Hadamard product.

Therefore, if the shapes and compositions of all modes can be determined from wide-area measurements, PFs can be directly computed. However, achieving full modal observability of the system is often not feasible due to insufficient phasor measurement units (PMUs) [30]. In fact, grid operators focus the real-time monitoring on selected dominant modes and thus do not have to pursue full modal observability. The MPFs are calculated as follows with subscripts *x* and *z* distinguishing the **x**- and **z**-spaces.

$$\mathbf{PF}_z = \mathbf{\Phi}_z \circ \mathbf{\Psi}_z^{\mathrm{T}} = (\mathbf{H}\mathbf{\Phi}_x) \circ (\mathbf{H}^{-\mathrm{T}}\mathbf{\Psi}_x^{\mathrm{T}}), \quad (9a)$$

$$\mathbf{PF}_x = \mathbf{\Phi}_x \circ \mathbf{H}^{\mathrm{T}} [\mathbf{PF}_z / (\mathbf{H}\mathbf{\Phi}_x)]. \quad (9b)$$

With partial modal observability, $\mathbf{\Phi}_z$ is only partially known, so the MPFs in the **x**-space cannot be obtained directly. Consequently, the MPFs in the **x**-space cannot be directly determined. Assuming that the transformation **H** from **x**-space to **z**-space has been obtained in Step 3, these relationships can be derived for elements of matrices $\mathbf{\Phi}_x$ and $\mathbf{\Psi}_x$:

$$[\mathbf{\Phi}_x]_{ij} = [\mathbf{H}^{-1}\mathbf{\Phi}_z]_{ij} = \sum_{k=1}^{N} m_{ik}\phi_{z,kj}, \quad (10a)$$

$$[\mathbf{\Psi}_x]_{ij} = [\mathbf{\Psi}_z\mathbf{H}]_{ij} = \sum_{k=1}^{N} h_{ki}\psi_{z,kj}. \quad (10b)$$

where $h_{ki}$ and $m_{ik}$ are the elements of the transformation matrix **H** and its inverse. Therefore, the PF for *i* state in *j* mode in **x**-space is

$$p_{x,ij} = \phi_{x,ij}\psi_{x,ji} = \sum_{k=1}^{N} m_{ik}\phi_{z,kj} \sum_{k=1}^{N} h_{ki}\psi_{z,jk}. \quad (11)$$

From the definition of PFs, there is

$$p_{x,ij} = \sum_{k=1}^{N} m_{ik}\phi_{z,kj} \sum_{k=1}^{N} h_{ki}\frac{p_{z,kj}}{\phi_{z,kj}}. \quad (12)$$

which associates MPFs in the **x**-space with corresponding mode shapes. In other words, if the right eigenvectors of certain modes in **z**-space are known, their corresponding MPFs in **x**-space can also be estimated after a linear transformation, even if the shapes and compositions of other modes are unknown.

### C. Error Estimation

If the linearized model of the system is available, an error index on the EPFs can be calculated according to (5):

$$e_i = \frac{1}{L}\sum_{l=1}^{L}\sum_{j=1, j\neq k}^{N}\psi_{jk}\phi_{ki}\frac{x_{j0}^{(l)}}{x_{k0}^{(l)}}. \quad (13)$$

Also, the following error index is introduced to evaluate the accuracy of the proposed MPFs compared to model-based PFs:

$$e_{2ij} = \left(\frac{\mathrm{PF}_j/\mathrm{PF}_i}{\mathrm{MPF}_j/\mathrm{MPF}_i} - 1\right)\times 100\%, \quad (14)$$

where the subscripts *i* and *j* are the generator numbers of interest.

## III. CASE STUDIES

This section first illustrates the proposed measurement-based approach for PF estimation on a two-area system, considering both full and partial observabilities. Penetration of inverter-based resources (IBRs) is also considered. Then, it is tested on the NPCC 48-machine system with both ideal and practical measurements.

### A. Two-area system study

A modified Kundur's two-area system [31] is used with generator inertias adjusted to present three distinct oscillatory modes at frequencies of 0.593Hz, 1.110Hz and 1.628Hz. The system includes generators 1 and 2 in Area 1, and generators 3 and 4 in Area 2. The 0.593 Hz mode is the inter-area mode, while the 1.110 Hz mode and 1.628 Hz mode are the intra-area modes of areas 1 and 2, respectively.

The mode compositions, mode shapes and PFs with four generators are computed from the model with respect to the inter-area mode. Fig. 4 presents a comparison of the results, revealing that generator 3 displays the most important role in the mode shape. Meanwhile, generator 1 contributes more significantly to the composition of the mode than any other generator. In summary, generator 1 exhibits the largest PF when considering both the mode shape and its composition. From this example, it can be concluded that PFs serve as reliable indicators for ranking generators according to their bidirectional associations with the mode.

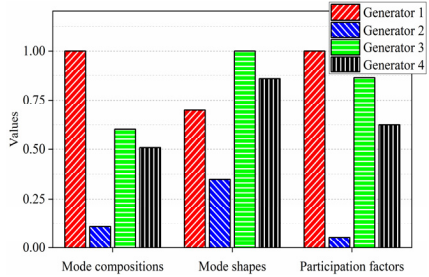

**Fig. 4.** Modal features on the interarea mode.

#### 1) Full observability.

Assuming that all four generators are monitored by PMUs, a database of measurements is created using simulation results on four rotor speeds under approximately 12 disturbances. Fig. 5 (left) displays measurements of the four speeds in two different 3D projections to depict the measurements in their 4D space. Fig. 5 (middle and bottom) depicts some symmetric states identified in the **x**-space, which approximately form a 3-parallelotope, and their transformation approximately forms a cuboid in the **z**-space.

TABLE II compares the MPFs calculated from the proposed PF estimation approach with the model-based PFs, which are found to be very similar. The errors in the estimation can be attributed to two factors: 1) the most symmetric initial states in **x**-space may not perfectly form a parallelotope (as evident in Fig. 5. 2) errors in the estimation due to Prony's analysis in **z**-space. Simulation results confirm that if the initial states selected from measurements form a perfectly symmetric parallelotope, the MPFs will be an exact match to the PFs. It is worth noting that when the black-box model is available and the disturbance can be designed, the MPF proposed approach is almost identical to the PFs.

In this section, a sub-space strategy for PF estimation is also proposed and tested on the same two-area system. The strategy involves identifying three 2-parallelotopes using six symmetric pairs of generators and calculating relative MPFs for each parallelotope. The MPFs are then normalized to obtain the MPFs for each generator. The results are shown in TABLE III, where the MPFs for the two local modes at 1.110 Hz and 1.628 Hz accurately match the PFs, while the MPFs for the inter-area mode at 0.593 Hz have larger errors but still reflect a similar ranking of the PFs. Therefore, the sub-space strategy is more suitable for PF estimation of local modes. However, for inter-area modes, the reduction in dimension needs to balance the trade-off between accuracy and computational complexity.

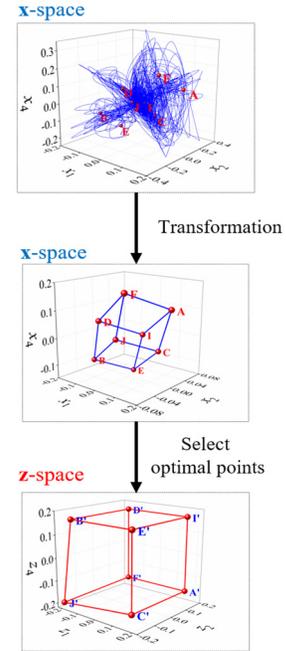

**Fig. 5.** The trajectory of response and selected initial states in state space.

TABLE II
THE PFs AND MPFs FOR FULL OBSERVABILITY STUDY

| Generator | 0.593 Hz | | 1.110 Hz | | 1.628 Hz | |
|---|---|---|---|---|---|---|
| | PFs | MPFs | PFs | MPFs | PFs | MPFs |
| 1 | 1.00 | 1.00 | $2.0\times10^{-3}$ | $2.6\times10^{-3}$ | 0.13 | 0.13 |
| 2 | 0.06 | 0.06 | $6.2\times10^{-4}$ | $5.0\times10^{-3}$ | 1.0 | 1.0 |
| 3 | 0.86 | 0.90 | 0.80 | 0.82 | $2.9\times10^{-3}$ | $2.5\times10^{-3}$ |
| 4 | 0.62 | 0.66 | 1.00 | 1.00 | $8.3\times10^{-3}$ | $9.2\times10^{-3}$ |

TABLE III
THE PFs AND MPFs BASED ON THE RATIO OF THE TWO-DIMENSION SYSTEM

| Generator | 0.593 Hz | | 1.110 Hz | | 1.628 Hz | |
|---|---|---|---|---|---|---|
| | PFs | MPFs | PFs | MPFs | PFs | MPFs |
| 1 | 1.00 | 1.00 | $2.0\times10^{-3}$ | $4.8\times10^{-4}$ | 0.13 | 0.12 |
| 2 | 0.06 | 0.17 | $6.2\times10^{-4}$ | $3.6\times10^{-3}$ | 1.00 | 1.00 |
| 3 | 0.86 | 1.06 | 0.80 | 0.62 | $2.9\times10^{-3}$ | $5.9\times10^{-3}$ |
| 4 | 0.62 | 0.97 | 1.00 | 1.00 | $8.3\times10^{-3}$ | $7.5\times10^{-3}$ |

### 2) Partial observability.

In this study, assume that not all generators are monitored by PMUs, and the state variables of unmeasured generators contain random changes following a uniform distribution. This research examines four distinct cases, with details provided in TABLE IV.

**Case 1**: Generators 1, 2 and 3 are monitored by PMUs, and the measurements of their speeds of the best symmetry are found. Generator 4, not monitored by a PMU, is assumed to have an initial speed change randomly from -0.02 to 0.02 rad/s.

**Case 2**: The same as Case 1 except that generator 3 also has a random initial speed as generator 4.

**Case 3**: The same as Case 1 except for an increased range of random initial speeds of generators 3 and 4.

**Case 4**: The same as Case 2 except for an increased range of random initial speeds of generators 3 and 4.

#### TABLE IV
#### THE DISTRIBUTION OF THE INITIAL STATE SET

| Case | generator 1 | generator 2 | generator 3 | generator 4 |
|------|-------------|-------------|-------------|-------------|
| 1 | Symmetric | Symmetric | Symmetric | [-0.02 0.02] |
| 2 | Symmetric | Symmetric | [-0.02 0.02] | [-0.02 0.02] |
| 3 | Symmetric | Symmetric | Symmetric | [-0.1 0.1] |
| 4 | Symmetric | Symmetric | [-0.1 0.1] | [-0.1 0.1] |

The study utilizes the error estimation presented in (14) to evaluate the performance of partial observability for generators 1 and 2, with the results outlined in TABLE V. A comparison between Cases 1 and 2, as well as Cases 3 and 4, indicates an increase in error as more generators lack PMU equipment. Additionally, a rise in the distribution range of generator 4 results in an increase in error, as evident in Cases 2 and 4. The error of the 1.110 Hz mode is significantly higher due to the critical roles played by generators 3 and 4, particularly in cases where the latter is not equipped with a PMU. Overall, the accuracy of the estimated MPF can be significantly enhanced if the initial speed variance of a generator without a PMU is small, particularly when it contributes significantly to the mode(s) of interest. Consequently, MPFs estimation demonstrates superior accuracy when more generators are equipped with PMUs.

#### TABLE V
#### THE ERROR INDEX OF GENERATOR 1 VS. 2

| Case | 0.593 Hz | 1.110 Hz | 1.628 Hz |
|------|----------|----------|----------|
| 1 | 0.78% | 5.66% | 0.35% |
| 2 | -1.51% | 7.23% | 1.02% |
| 3 | -0.82% | 12.67% | 1.04% |
| 4 | -1.65% | 48.13% | 1.97% |

### 3) Considering penetration of IBRs

Another case study was conducted on the two-area system, replacing Generators 2, 3, and 4 by IBRs. The detailed model of the IBR can be found in [32] and [33]. Twelve disturbances were considered, and the results are shown in TABLE VI.

Due to the penetration of IBRs, the frequencies of the modes become much faster. The inter-area mode is around 1.05 Hz, while the local modes are around 1.95 Hz and 2.22 Hz. The MPFs are closely aligned with the PFs, although Gen 3 and Gen 4 exhibit relatively larger errors. This discrepancy arises because the two IBRs are located far from the synchronous generators, and the coherence between these two generators is more pronounced. The differences are notable for Gen 1 and 2, making it easier to distinguish between them.

#### TABLE VI
#### THE PFS AND MPFS FOR FULL OBSERVABILITY STUDY

| Generator | 1.05 Hz | | 1.95Hz | | 2.22 Hz | |
|-----------|---------|------|--------|------|---------|------|
| | PFs | MPFs | PFs | MPFs | PFs | MPFs |
| 1 | 0.83 | 0.82 | 0.51 | 0.49 | $2.5 \times 10^{-3}$ | $1.9 \times 10^{-2}$ |
| 2 | 0.51 | 0.37 | 1.00 | 1.00 | $3.6 \times 10^{-3}$ | $8.6 \times 10^{-2}$ |
| 3 | 1.00 | 1.00 | $1.6 \times 10^{-2}$ | 0.11 | 0.57 | 0.72 |
| 4 | 0.60 | 0.48 | $4.3 \times 10^{-3}$ | 0.24 | 1.00 | 1.00 |

### B. NPCC System

Next, the proposed approach's performance is tested on a much larger NPCC 140-bus 48-machine system, as illustrated in Fig. 6. Two distinct measurement datasets will be generated: random initial states and faults. The testing will concentrate on five generators (21, 24, 26, 27, and 78) with the largest PFs of the 0.6 Hz inter-area mode.

### 1) Test on faults.

This test simulates 50 three-phase faults on the lines near one end in the New England region of the system, as shown in Fig. 6. Each fault lasts 50 ms before being cleared, and no generator or line is tripped after the fault. A total of 50 fault simulations will be executed to generate the measurement dataset, which will then be used to estimate MPFs. The outcomes of the test will be presented in TABLE VII.

From the results, only the MPFs of generators 78 and 26 are close to their true values. The MPFs of the remaining three generators are inaccurate due to the strong coherence observed among the last four generators in the oscillation. The highest mutual coherence of 0.98 is observed between the columns formed by generators 24 and 21, with condition number 249. Consequently, the sampling matrix is ill-conditioned, significantly amplifying the input error. Detailed proof of the relationship between error and coherence can be found in the appendix.

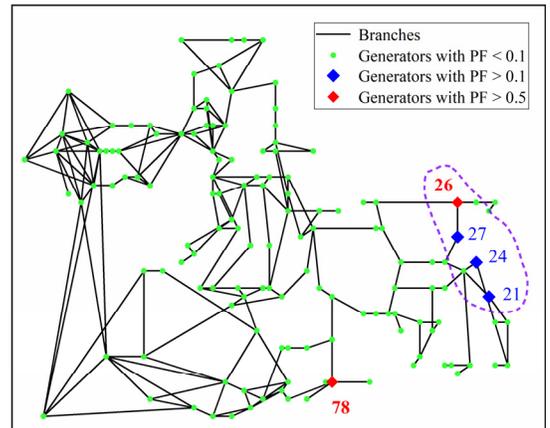

**Fig. 6.** Topology of the NPCC 140-bus system

| Generator | 78 | 26 | 24 | 21 | 27 |
|---|---|---|---|---|---|
| (True) PFs | 1.00 | 0.54 | 0.43 | 0.18 | 0.10 |
| MPFs | 0.99 | 0.61 | 1.00 | 0.88 | 0.31 |

*4) Test on the dataset of random initial states.*

Based on the findings from the two-area system study, it is established that the performance of the proposed approach is significantly influenced by the distribution of initial states within the measurement dataset. To validate this observation on the NPCC system, a measurement dataset will be generated by conducting simulations of scenarios starting from randomly selected initial states within the state space. The selected initial states will be sufficiently disturbed within the state space. A total of 50 scenarios will be generated, each simulating for 20 seconds, resulting in the measurement dataset.

The MPFs estimated using data from 10, 20, 30, 40, and all 50 simulations are presented in TABLE VIII. The results indicate that utilizing data from 20 simulations can correctly identify the generator with the largest PF. Increasing the number of simulations to 40 results in the correct ranking of PFs. When all 50 scenarios are applied, the MPFs closely match the true PFs, thus demonstrating the effectiveness of the measurement-based PF approach for the large NPCC system.

TABLE VIII
THE PFs BASED ON RANDOM INITIAL STATES

| Generator | True PFs | MPFs based on several scenarios | | | | |
|---|---|---|---|---|---|---|
| | | 10 | 20 | 30 | 40 | 50 |
| 78 | 1.00 | 0.95 | 1.00 | 1.00 | 1.00 | 1.00 |
| 26 | 0.54 | 0.99 | 0.50 | 0.61 | 0.60 | 0.55 |
| 24 | 0.43 | 1.00 | 0.42 | 0.75 | 0.36 | 0.40 |
| 21 | 0.18 | 0.13 | 0.17 | 0.50 | 0.25 | 0.17 |
| 27 | 0.10 | 0.12 | 0.25 | 0.72 | 0.15 | 0.16 |

## IV. CONCLUSION

This paper has developed a measurement-based approach for PF estimation. The computed MPFs are the best approximations of the model-based theoretical PFs by means of a linear coordinate transformation that relaxes the symmetric condition for calculating PFs from responses of the system. The efficacy of the proposed approach is demonstrated through its applications to both small- and large-scale power system models with discussions on its error.

## APPENDIX

### A. The invariant of transformation

**Lemma 1** The transformation from a parallelotope to a hyperrectangle of the same dimension is invariant under a translation of coordinates.

*Proof*: Assume the **x**-space is a vector space with basis $\{\mathbf{u}_1, \ldots, \mathbf{u}_N\}$. Since **z**-space is a vector space transformed from the **x**-space by transformation **H**. Thus, $\{\mathbf{v}_1, \ldots, \mathbf{v}_N\}$ is a set of basis in **z**-space with $\mathbf{v}_j = \mathbf{H}\mathbf{u}_j$ ($j=1, \ldots, N$). For a random translation $\mathbf{s}_{\mathbf{x},0}$ with coordinate $[s_1, \ldots, s_N]^\mathrm{T}$ in **x**-space, it can be noticed that

$$\mathbf{H}\mathbf{s}_{x,0} = \mathbf{H}\sum_{j=1}^{N} s_j \mathbf{u}_j = \sum_{j=1}^{N} s_j \mathbf{H}\mathbf{u}_j = \sum_{j=1}^{N} s_j \mathbf{v}_j = \mathbf{s}_{z,0}. \quad (14)$$

Hence, for a random initial state in **x**-space,

$$\mathbf{H}(\mathbf{x}_i - \mathbf{s}_{x,0}) = \mathbf{H}\mathbf{x}_i - \mathbf{H}\mathbf{s}_{x,0} = \mathbf{z}_i - \mathbf{s}_{z,0}. \quad (15)$$

Therefore, for any $\mathbf{x}_i$ with translation $\mathbf{s}_{\mathbf{x},0}$ in **x**-space will be transformed by **H** to the initial state $\mathbf{z}_i = \mathbf{H}\mathbf{x}_i$ with the translation $\mathbf{s}_{z,0} = \mathbf{H}\mathbf{s}_{x,0}$ in **z**-space. In other words, **H** is invariant under a translation of coordinates.

First, translate the origin to one of the $N$+1 vertices, say $\mathbf{x}_0$. Vectors $\mathbf{x}_1-\mathbf{x}_0$, …, $\mathbf{x}_N-\mathbf{x}_0$ together define a convex polyhedral cone as well as a set of basis vectors [34]. Thus, for any of these vectors $\mathbf{x}_i-\mathbf{x}_0$, there is

$$\mathbf{x}_i - \mathbf{x}_0 = \sum_{j=1}^{N} a_j(\mathbf{x}_j - \mathbf{x}_0) \quad a_j = \{0,1\}. \quad (16)$$

It is easy to notice that if let $\mathbf{z}_i - \mathbf{z}_0 = [a_1, \ldots, a_N]^{-1}$, there is

$$\mathbf{H}^{-1}(\mathbf{z}_i - \mathbf{z}_0) = \sum_{j=1}^{N} a_j(\mathbf{x}_j - \mathbf{x}_0) = \mathbf{x}_i - \mathbf{x}_0. \quad (17)$$

Because $a_j$ takes only 0 or 1 for any $j$, $\mathbf{z}_i - \mathbf{z}_0$ is actually the vertex on the hyperrectangle. Thus, **H** maps vertices of the $N$-parallelotope in **x**-space after the translation to vertices of a hyperrectangle in **z**-space. Also, since **H** is linear and invertible with a zero kernel, it is injective. Thus, all $2^N$ vertices on the $N$-parallelotope can be transformed by **H** to the $2^N$ unique vertices on the hyperrectangle.

Besides, from *Lemma 1*, the translation to $\mathbf{x}_0$ does not change **H**, so H is the desired transformation.

∎

### B. The relationship between coherence and error

The proposed approach can be represented as solving a linear equation in the following form:

$$\mathbf{S}\boldsymbol{\Psi} = \mathbf{B}, \quad (18)$$

where **S** is the sampling matrix of measurements, with each row representing the rotor speed data of all generators (at different columns) at a particular time instant of the measuring window. **Ψ** is the mode composition that needs to be computed, while **B** represents the excitation energy calculated from the measurement. Given a sufficiently high sampling frequency of measurements, **S** generally has significantly more rows than columns, making (19) an overdetermined system.

However, if the rotor speeds of two generators (say generators $i$ and $j$) are almost proportional to each other, the columns $i$ and $j$ of S become coherent, meaning that these two vectors are nearly linearly dependent. This results in a large condition number of **S** and a considerable error in the solution of (19). The coherency index $\gamma$ of **S** is defined as

$$\gamma = \max_{1 \le \alpha \le N, 1 \le \beta \le N} \left\| \mathbf{s}_\alpha^\mathrm{T} \mathbf{s}_\beta \right\| / \left( \left\| \mathbf{s}_\alpha \right\| \left\| \mathbf{s}_\beta \right\| \right) = \left\| \mathbf{s}_i^\mathrm{T} \mathbf{s}_j \right\| / \left( \left\| \mathbf{s}_i \right\| \left\| \mathbf{s}_j \right\| \right), \quad (19)$$

where $\mathbf{s}_\alpha$ and $\mathbf{s}_\beta$ are arbitrary two-column vectors of the matrix **S**, and $\mathbf{s}_i$ and $\mathbf{s}_j$ are two-column vectors that have the smallest angle in between. The condition number of a matrix is typically defined by using its nearest lower-rank matrix [35].

$$\mathrm{cond}(\mathbf{S}) = \max \left\| \mathbf{S} \right\|_2 / \left\| \mathbf{S} - \hat{\mathbf{S}} \right\|_2, \ \mathrm{cond}(\mathbf{S}) = \max \left\| \mathbf{S} \right\|_2 / \left\| \mathbf{S} - \hat{\mathbf{S}} \right\|_2, \quad (20)$$

The 2-norm of a matrix is the square root of the sum of all its elements squared. Consider some matrix $\hat{\mathbf{S}}$ that is the same as $\mathbf{S}$ except that column $j$ equals its column $i$. Namely,

$$\hat{\mathbf{S}} = [\mathbf{s}_1, \quad \ldots, \quad \mathbf{s}_{j-1}, \quad \mathbf{s}_i, \quad \mathbf{s}_{j+1}, \quad \ldots, \quad \mathbf{s}_N]. \tag{21}$$

It has a lower rank reduced by 1. From (21),

$$\text{cond}(\mathbf{S}) \geq \frac{\|\mathbf{s}\|_2}{\|\mathbf{S} - \hat{\mathbf{S}}\|_2} = \frac{\|\mathbf{s}\|_2}{\|\mathbf{s}_i - \mathbf{s}_j\|_2} = \frac{\|\mathbf{s}\|_2}{\sqrt{|\mathbf{s}_i|^2 + |\mathbf{s}_j|^2 - 2|\mathbf{s}_i||\mathbf{s}_j|\gamma}}. \tag{22}$$

Thus when $\gamma \to 1$, there is

$$|\mathbf{s}_i|^2 + |\mathbf{s}_j|^2 - 2|\mathbf{s}_i||\mathbf{s}_j|\gamma \to 0 \quad \text{cond}(\mathbf{S}) \to \infty, \tag{23}$$

which means that the condition number will become extremely large. Since the sampling matrix is ill-conditioned, errors will become inevitable.